\documentclass{elsarticle}
\usepackage{graphicx}%
\usepackage{multirow}%
\usepackage{amsmath,amssymb,amsfonts}%
\usepackage{amsthm}%
\usepackage{mathrsfs}%
\usepackage[title]{appendix}%
\usepackage{xcolor}%
\usepackage{textcomp}%
\usepackage{manyfoot}%
\usepackage{booktabs}%
\usepackage{algorithm}%
\usepackage{algorithmicx}%
\usepackage{algpseudocode}%
\usepackage{listings}%
\usepackage{verbatim}
\usepackage{enumitem}
 \usepackage{comment}
 \usepackage{amsmath}

\journal{}
\begin{document}

\begin{frontmatter}

\title{Exact Solution of Granovetter's Threshold Model for a Finite Population }

\author[mymainaddress]{Jos\'e F.  Fontanari}

\address[mymainaddress]{Instituto de F\'{\i}sica de S\~ao Carlos,  Universidade de S\~ao Paulo,  13566-590 S\~ao Carlos,  S\~ao Paulo,  Brazil}

\begin{abstract}
The Granovetter threshold model formalizes collective behavior by assuming that individual agents face a binary decision to join a movement, doing so only when the number of already active participants reaches or exceeds an intrinsic, personal threshold. In this work, we derive an exact analytical expression for the probability that a cascade halts with precisely $k$ active agents in a finite population of size $N$ triggered by a single initial instigator, and use this result to obtain the scaling corrections that govern the system near its critical boundaries. By parameterizing individual threshold heterogeneity via a Beta distribution with shape parameters $\alpha$ and $\beta$, we map how these micro-level predispositions aggregate into macro-level collective outcomes. Here, a small $\alpha$ represents a high proportion of low-threshold, highly susceptible agents, while a small $\beta$ marks a significant density of high-threshold, conservative individuals. In the infinite-population limit, a phase transition occurs at the critical parameter $\alpha = 1$, which separates an inactive phase from a regime of widespread mobilization. For a power threshold distribution ($\beta = 1$), the system exhibits a discontinuous, first-order phase transition where the active fraction jumps abruptly from 0 to 1, and the finite-size critical scaling window contracts as $N^{-1/2}$. In stark contrast, when the population features a persistent density of high-threshold agents ($\beta < 1$), the system undergoes an infinite-order phase transition characterized by an exceptionally smooth, continuous onset of collective activity, causing the finite-size critical region to contract at a drastically slower rate proportional to $(\ln N)^{-1}$. These analytical findings establish a mathematical benchmark for finite-size effects in behavioral cascades.
\end{abstract}

\end{frontmatter}


\section{Introduction}\label{sec:Intro}

The formalization of collective movements---ranging from residential segregation, riots, and strikes to the diffusion of rumors and technological innovations---rests on the conceptual foundation of threshold models \cite{Granovetter_1978}. The intellectual lineage of this framework traces back to the pioneering work of Thomas Schelling in the 1970s, who introduced the first threshold-like spatial model to demonstrate how localized, micro-level individual preferences regarding neighborhood composition could aggregate into stark, macro-level patterns of racial segregation \cite{Schelling_1971}. Schelling's insights established that social systems often exhibit non-linear tipping points, where minor shifts in a population's composition can trigger abrupt, large-scale structural transformations. The underlying dynamics of these phenomena typically unfold as a self-reinforcing chain reaction, widely conceptualized in the modern literature as a behavioral cascade \cite{Bikhchandani_1992,Macy_2020}.

In his seminal work, Mark Granovetter shifted the focus from Schelling's spatial neighbor-interaction dynamics to a non-spatial, fully connected social framework, mathematically formalizing the cascade mechanism for binary decisions within an interacting collective \cite{Granovetter_1978}. In Granovetter's model, a population of $N$ interacting agents is considered, where individuals choose between non-participation or participation---defined as entering an active state---in a collective action. The dynamics proceed in discrete time steps under a synchronous update rule, where all agents simultaneously re-evaluate their state at each interval. The decision of a given agent $i$ to transition into the active state is governed exclusively by an intrinsic, individual threshold $T_i$, which represents the minimum number of other active population members required for agent $i$ to join the movement \cite{Granovetter_1978}.

In general, subsequent extensions of Granovetter's framework have sought to introduce greater structural realism by relaxing the assumption of a fully connected population. A substantial body of literature has explored threshold dynamics on complex networks, mapping how topologies such as small-world, scale-free, or clustered networks accelerate or hinder global cascades \cite{Watts_2002, Centola_2005}. Crucially, many of these models incorporate heterogeneous connection weights to reflect the varying degrees of social influence between individuals. This friendship effect---where the actions of close acquaintances carry significantly more weight than those of strangers---was already explicitly emphasized by Granovetter in his original 1978 article, where he noted that the structure of social ties could fundamentally alter or even overwhelm individual preferences \cite{Granovetter_1978}.

While  network-based models offer invaluable insights into empirical social systems, their added complexity often precludes closed-form analytical treatments, rendering them heavily reliant on numerical simulations. The objective of the present work is not to propose a more realistic, topologically complex model; a vast literature has already advanced extensively along those trajectories \cite{Watts_2002, Centola_2005,Centola_2007,Akhmetzhanov_2013,Wiedermann_2020,Min_2023,Abella_2024}.  Instead, our goal is to return to the  fundamental principles of the original formulation  and offer an exact analytical solution to the Granovetter model, where all agents interact globally with one another under a regime of homogeneous connection weights. By solving this baseline case exactly, we establish a  mathematical benchmark that clarifies the fundamental mathematical properties of behavioral cascades.

To operationalize the heterogeneity of individual predispositions within a continuous framework, we model an agent's intrinsic willingness to participate as a continuous random variable $x \in [0, 1]$. To map these continuous dispositions onto a finite population of size $N$, the discrete threshold $T_i \in \{1, \dots, N\}$ is set via $T_i = 1 + \lfloor N x \rfloor$. To achieve a flexible representation of these profiles, we draw $x$ from a Beta distribution, which is uniquely suited for this purpose as its support is strictly bounded on the unit interval. Explicitly, the probability density function of the threshold distribution is given by
\begin{equation}\label{eq:Beta}
f(x; \alpha, \beta) = \frac{x^{\alpha-1}(1-x)^{\beta-1}}{B(\alpha, \beta)}, \quad x \in [0, 1]
\end{equation}
where $\alpha > 0$ and $\beta > 0$ are the continuous shape parameters that dictate the distribution's profile, and $B(\alpha, \beta)$ is the complete Beta function \cite{Abramowitz_1972}. The corresponding cumulative distribution function, which represents the total fraction of the population with predispositions less than or equal to $x$, is denoted by $F(x) = \int_0^x f(u; \alpha, \beta) \, du$.  By varying $\alpha$ and $\beta$, this formulation allows us to map the entire spectrum of population profiles---from highly radicalized societies with low thresholds ($\alpha < 1, \beta > 1$) to highly resistant populations ($\alpha > 1, \beta < 1$), as well as the classic uniform baseline when $\alpha = \beta = 1$.

The main contribution of this work is the study of the finite-population Granovetter model under the initial condition of a single instigator, i.e., an agent with threshold $T_i=0$. We focus mainly on the characterization of critical phenomena where minor changes in the threshold distribution lead to massively distinct macroscopic outcomes. In particular, we show that a single instigator is insufficient to trigger a large-scale collective action when both $\alpha \ge 1$ and $\beta \ge 1$. In this highly moderate or conformist regime, the population lacks the necessary density of low-threshold individuals to catch the initial spark, causing the cascade to choke at its very inception and leaving the collective movement trapped in an inactive phase. However, large-scale mobilization becomes dynamically accessible when $\beta \le 1$, occurring specifically within the regime where $\alpha < 1$.

In the limit of an infinite population, a  phase transition occurs exactly at the critical point $\alpha=1$, which separates a regime where a macroscopic fraction of the population becomes active from a regime where the active fraction vanishes. By using finite-size scaling methods \cite{Privman_1990,Landau_2000} alongside exact large-deviation analyses \cite{Touchette_2009}, we derive the precise scaling corrections that govern the system for finite $N$ in the vicinity of this critical point. Our results reveal that for the power threshold distribution ($\beta = 1$), the infinite-system transition is  discontinuous, characterized by an abrupt jump in the steady-state active fraction from $1$ when $\alpha < 1$ to $0$ as soon as $\alpha > 1$.  In finite populations, this first-order transition manifests as a steep, rapid rise in the active fraction within a critical region that shrinks as $N^{-1/2}$. In stark contrast, when $\beta < 1$, the system undergoes an infinite-order phase transition where the proportion of active agents increases continuously from zero, and the finite-size critical scaling window contracts at a drastically slower rate proportional to $(\ln N)^{-1}$.

The rest of this paper is organized as follows. For the sake of completeness, Section \ref{sec:inf} offers a brief overview of the well-known infinite-population limit of Granovetter's model for a general threshold distribution. Section \ref{sec:fin} introduces our key analytical contribution, deriving the exact expression for the probability that a cascade halts with precisely $k$ active agents in a population of size $N$. These exact finite-size results serve as the basis for calculating the precise asymptotic behavior within the critical regions via a large-deviation analysis, the comprehensive mathematical proofs of which are detailed in the four Appendices of this paper. We then apply this exact framework to specific population profiles: Section \ref{sec:PF} examines both the infinite and finite-population scenarios under a power threshold distribution ($\beta=1$), while Section \ref{sec:B} addresses the more general Beta distribution regime where $\beta<1$. Finally, Section \ref{sec:conc} presents our concluding remarks.

\section{The Infinite Population Limit}\label{sec:inf}

In the infinite population limit ($N \to \infty$), the model admits a straightforward analytical description \cite{Granovetter_1978,Granovetter_1983}. Since the intrinsic thresholds are drawn independently, the Law of Large Numbers \cite{Feller_1968} ensures that the empirical fraction of the population with a normalized threshold $T/N \le x$ converges deterministically to the cumulative probability $F(x)$.

Let $\rho_t$ denote the macroscopic fraction of active individuals at a given cascade step $t$. The cascade operates synchronously: the agents who will be active at the subsequent step $t+1$ are precisely those whose thresholds are satisfied by the current active fraction $\rho_t$. Therefore, the macroscopic propagation of the cascade is perfectly described by the deterministic recurrence mapping
\begin{equation}\label{rec_t}
\rho_{t+1} = F(\rho_t).
\end{equation}
The cascade evolves until it reaches a stationary state, which corresponds to the stable fixed points of this mapping. In the language of statistical physics, the steady-state fraction $\rho_\infty$ acts as the order parameter of the system and is determined by the self-consistency equation
\begin{equation}\label{rec_*}
\rho_\infty = F(\rho_\infty).
\end{equation}
The specific shape of the distribution $F(x)$ encapsulates the societal heterogeneity and polarization. Since the thresholds are strictly bounded within the unit interval, we have $F(0)=0$ and $F(1)=1$; consequently, $\rho_\infty=0$ (microscopic cascade) and $\rho_\infty=1$ (global cascade) are always solutions of this equation. Depending on the choice of the cumulative distribution $F(x)$, additional solutions may appear between these extremes \cite{Granovetter_1978,Granovetter_1983}.
 In particular, for the Beta distribution adopted in this work, the system admits at most one intermediate solution ($0 < \rho_\infty < 1$). Consequently, the cascade dynamics undergoes a phase transition as the parameters $\alpha$ and $\beta$ of the Beta distribution vary and alter the stability of the solutions of Equation (\ref{rec_*}).
 
 However, for any finite system size $N$, this deterministic picture is incomplete. The macroscopic mapping breaks down because the cascade dynamics become intrinsically stochastic, driven by finite-size sampling fluctuations in the  realization of individual thresholds. To properly understand the phase transition behavior and compute the exact distribution of the final cascade size, a thorough finite-population analysis is required.

\section{Analytical Formulation for Finite Populations}\label{sec:fin}

We aim to find $P_N(k)$, defined as the  probability that the cascade process halts with exactly $k$ active individuals in a total population of size $N$. To trigger the dynamics, we assume the system is initialized with a single instigator (an agent with threshold $T=0$).

For the cascade to freeze at an exact final size $k$, the population splits into two independent, non-overlapping groups whose roles must satisfy two distinct probabilistic conditions simultaneously:

\begin{enumerate}[label=(\roman*)]
\item A specific subset of $k-1$ individuals, together with the initial instigator, must successfully sustain an unbroken, step-by-step recruitment process until all $k$ of them are active. We define $Q_k$ as the joint probability that this isolated cluster of $k$ individuals successfully completes this sequential activation from size $1$ up to $k$ without choking prematurely at any intermediate size $j < k$.

\item The remaining $N-k$ individuals in the population must strictly refuse to join the movement when it reaches size $k$. This structural barrier guarantees that their individual thresholds are strictly greater than $k$, thereby preventing any further propagation of the cascade. The probability of this occurring for each of these individuals is $1 - F(k/N)$.
\end{enumerate}
Combining these two conditions using a binomial distribution---since we are selecting $k-1$ specific participants out of the $N-1$ available individuals---we obtain the  probability mass function for the final cascade size
\begin{equation}\label{eq:PN}
P_N(k) = \binom{N-1}{k-1} Q_k \left[ 1 - F\left(\frac{k}{N}\right) \right]^{N-k}.
\end{equation}
The propagation factor $Q_k$ can be determined  by invoking a probability conservation argument.  Suppose we isolate a subpopulation of size $k$ that contains the instigator and $k-1$ individuals known to have thresholds $\le k$. If we run the cascade dynamics strictly within this isolated subpopulation,  the cascade must terminate at some specific size $j \in \{1, 2, \dots, k\}$. Consequently, the sum of the probabilities of stopping at any size $j$ up to $k$ must equal 1.

The probability that the cascade stops exactly at an intermediate size $j$ within this subpopulation is given by the probability that it successfully reaches size $j$ (which is $Q_j$) multiplied by the probability that the remaining $k-j$ individuals are not activated by $j$ people. Since the thresholds are evaluated against the global population scale $N$, the non-activation probability is $\left[ 1 - F(j/N) \right]$. Thus, the normalization condition is expressed as
\begin{equation}\label{eq:nQk}
\sum_{j=1}^{k} \binom{k-1}{j-1} Q_j \left[ 1 - F\left(\frac{j}{N}\right) \right]^{k-j} = 1.
\end{equation}
By isolating the last term of the sum ($j=k$), we obtain a  recursive relation for $Q_k$
\begin{equation}\label{eq:Qk}
Q_k = 1 - \sum_{j=1}^{k-1} \binom{k-1}{j-1} Q_j \left[ 1 - F\left(\frac{j}{N}\right) \right]^{k-j},
\end{equation}
with the base case $Q_1 = 1$, representing the certainty of the instigator's initial activation.

To gain physical intuition into the nature of the propagation factor, it is instructive to evaluate Equation (\ref{eq:Qk}) explicitly for the first few cascade sizes. For $k=2$, the recurrence relation yields
\begin{equation}
Q_2 = 1 - \left[1 - F\left(\frac{1}{N}\right)\right] = F\left(\frac{1}{N}\right).
\end{equation}
This result is intuitive: for the cascade to successfully grow from the initial instigator ($j=1$) to size $2$, the single chosen individual must be activated by the instigator alone, which occurs with probability $F(1/N)$. Moving to $k=3$, the expression becomes
\begin{equation}
Q_3 = 1 - \left[1 - F\left(\frac{1}{N}\right)\right]^2 - 2 Q_2 \left[1 - F\left(\frac{2}{N}\right)\right].
\end{equation}
%
%
%
We note that $Q_k$ acts as a cumulative survival probability of the process. In the case of $Q_3$, the system avoids dying out at size $j=1$ (the first subtracted term, where neither of the two available individuals can be triggered by the single instigator) and also avoids stalling at size $j=2$ (the second subtracted term, where the cascade successfully reaches size 2 but the last remaining individual refuses to join even when exposed to two active peers). Therefore, $Q_k$ effectively filters out all micro-histories where the collective movement chokes prematurely due to local gaps in the threshold distribution, ensuring a continuous, unbroken path of recruitment up to size $k$.

Using Equations (\ref{eq:PN}) and (\ref{eq:Qk}), the complete distribution $P_N(k)$ can be computed recursively for any arbitrary continuous cumulative distribution function $F(x)$. In this finite-size stochastic framework, the macroscopic order parameter---representing the expected final fraction of active individuals in the population---is given by the expectation value
\begin{equation}
\rho_\infty = \frac{1}{N} \sum_{k=1}^N k P_N(k). 
\label{eq:rho_inf_sum}
\end{equation}
In the infinite population  limit ($N \to \infty$), this expectation value converges precisely to the deterministic fixed point governed by the self-consistency condition (\ref{rec_*})  discussed in Section \ref{sec:inf}. In fact, in Appendix \ref{appA} we show that the propagation factor scales as
\begin{equation}
Q_k \approx \left[ F\left(\frac{k}{N}\right) \right]^k
\end{equation}
in the infinite population limit.  This asymptotic form offers a simplification of the cascade's micro-dynamics. It implies that in a sufficiently large population, the sequential, step-by-step history of the recruitment process becomes irrelevant. The term $[F(k/N)]^k$ simply represents the joint probability that $k$ independent individuals possess individual thresholds accommodating an active fraction of size $k/N$. Therefore, the asymptotic convergence of $Q_k$ guarantees that if a cluster of $k$ individuals capable of sustaining a cascade of that size exists, the internal propagation will never choke or stall at any intermediate size $j < k$. The macroscopic growth becomes deterministic, clearing the path to directly evaluate specific societal threshold distributions.

Substituting the asymptotic form of $Q_k$ into Equation (\ref{eq:PN}) reduces the probability mass function to a binomial distribution with success probability $p = F(k/N)$. As the population size $N$ grows infinitely large, the variance of the active fraction $k/N$ scales as $1/N$ and its fluctuations vanish, locking the system onto its expected value. Consequently, the expectation value in Equation (\ref{eq:rho_inf_sum}) collapses directly to the deterministic self-consistency condition given by Equation (\ref{rec_*}), demonstrating that the stochastic finite-population dynamics converge smoothly to the deterministic  framework described in Section \ref{sec:inf}.

Having established the basic theory for both finite and infinite populations, we now transition to investigating specific threshold distributions to extract concrete physical insights and analyze the resulting phase transitions. It is worth noting that while the recursive framework given by Equations~(\ref{eq:PN}) and (\ref{eq:Qk}) is formally exact, it becomes computationally intractable for populations larger than $N \approx 100$ due to the catastrophic propagation of numerical round-off errors inherent to nested combinatorial sums. To overcome this limitation and access the large-$N$ regime, we employ stochastic Monte Carlo simulations throughout this work. We have thoroughly verified that for small system sizes, where numerical errors remain  controlled, the Monte Carlo data perfectly reproduce the exact probabilities dictated by our analytical expressions. Crucially, however, the primary value of these exact equations is not numerical evaluation, but rather mathematical analysis; they serve as the indispensable starting point for the large-deviation analysis used to derive the exact asymptotic behavior of the cascades.

\section{The Power Function Threshold Distribution}\label{sec:PF}

A natural and highly flexible choice that fits the bounded domain $x \in [0,1]$ is the power function distribution, which arises as a special case of the Beta distribution when setting the second shape parameter $\beta = 1$,
\begin{equation}
F(x) = x^\alpha, \quad \text{with } \alpha > 0.
\label{eq:power_dist}
\end{equation}
The parameter $\alpha$ acts as a societal susceptibility index. When $\alpha = 1$, Equation (\ref{eq:power_dist}) simplifies to the standard uniform distribution, representing a homogeneous spread of thresholds. Surprisingly, as we  demonstrate in Appendix \ref{appB}, this baseline case allows for an exact, closed-form solution of the recursion equation (\ref{eq:Qk}) for the the propagation factor $Q_k$.

Tuning $\alpha > 1$ models a highly resistant, conservative population dominated by high thresholds, whereas $0 < \alpha < 1$ describes a volatile, easily triggered society filled with low-threshold agents. Notably, this distribution has proven highly effective in analyzing generalizations of Granovetter's model, such as frameworks accounting for the competition of opposing social movements over a population \cite{Ishikawa_2026}.

\subsection{Deterministic Analysis}

In the infinite population limit ($N \to \infty$), the final macroscopic fraction of active individuals is dictated by the deterministic self-consistency condition (\ref{rec_*}). Substituting the power function distribution from Equation (\ref{eq:power_dist}), we obtain the governing fixed-point equation
\begin{equation}
\rho_\infty = \rho_\infty^\alpha.
\end{equation}
Restricting our analysis to the physical domain $\rho_\infty \in [0, 1]$, this equation admits only two isolated fixed points for any $\alpha \neq 1$: the state of complete cascade failure ($\rho_\infty = 0$) and the state of a global, system-wide cascade ($\rho_\infty = 1$). The marginal case $\alpha = 1$ constitutes an exception, where the expression degenerates into an identity and every fraction within the continuous interval $[0, 1]$ becomes a fixed point.

To determine whether these macroscopic states are physically attainable, we must evaluate their dynamical stability. A fixed point $\rho_\infty$ is locally stable against infinitesimal perturbations  if the derivative of the driving function satisfies $|F'(\rho_\infty)| < 1$, and unstable if $|F'(\rho_\infty)| > 1$ \cite{Britton_2003}. For our threshold distribution, the derivative is given by $F'(x) = \alpha x^{\alpha - 1} \geq 0$, meaning that the stability of the system splits precisely into three distinct sociological regimes dictated by the susceptibility index $\alpha$.

In a conservative society ($\alpha > 1$), where the population is dominated by high-threshold individuals, evaluating the derivative at the origin yields $F'(0) = 0 < 1$, rendering the zero-cascade state ($\rho_\infty = 0$) a locally stable attractor. Conversely, at the fully active state, $F'(1) = \alpha > 1$, which means the global cascade ($\rho_\infty = 1$) is dynamically unstable. Sociologically, this implies that  macroscopic cascades cannot emerge from an infinitesimally small local fluctuation; the system possesses a strong social inertia, and any microscopic cascade attempt will inevitably dampen and collapse back to zero.

This stability profile perfectly inverts in the case of a volatile society ($0 < \alpha < 1$). When the population is saturated with low-threshold agents, the derivative at the origin diverges ($F'(0) = \infty$), making the zero-cascade state violently unstable. Meanwhile, at the global state, we find $F'(1) = \alpha < 1$, establishing $\rho_\infty = 1$ as a  stable global attractor. In this highly radicalized regime, the society is so primed for collective action that even a microscopic initial disturbance will autonomously amplify, chain-reacting until it inevitably engulfs the entire population.

Finally, at the exact boundary between volatility and conservatism, the system settles into the uniform baseline ($\alpha = 1$). Here, the derivative simplifies to $F'(x) = 1$ everywhere, causing the self-consistency condition to degenerate into the identity $\rho_\infty = \rho_\infty$. In this special, marginal case, every fraction in the continuous interval $[0,1]$ acts as a marginally stable fixed point. Without internal amplification or macroscopic dampening, the final size of the cascade is strictly dictated by the exact size of the initial triggering seed.

This purely deterministic analysis provides a clean macroscopic phase diagram of the society. However, to capture the microscopic, step-by-step probability of these cascades successfully surviving the vulnerable early stages of propagation, we must turn to the exact finite-size stochastic formulation.

\subsection{Finite Population Analysis}

While the deterministic limit successfully identifies the macroscopic attractors of the society, the actual emergence of a global cascade depends entirely on its ability to survive the microscopic, highly stochastic early stages of propagation. To investigate these finite-size dynamics, we initially explore the system through explicit Monte Carlo simulations. 

We consider a finite population of size $N$ initialized with exactly one active instigator (the initial seed). The thresholds of the remaining $N-1$ observers are independently drawn from the power function cumulative distribution given by Equation~(\ref{eq:power_dist}). Because the early stages of the cascade are extremely sensitive to the specific micro-configuration of thresholds, we performed extensive ensemble averaging to obtain reliable statistics. For each chosen system size and value of $\alpha$, the expected final active fraction $\rho_\infty$ was averaged over $10^5$ independent realizations, where each run represents a completely fresh random assignment of individual thresholds.

\begin{figure}[th]
\centering
 \includegraphics[width=1\columnwidth]{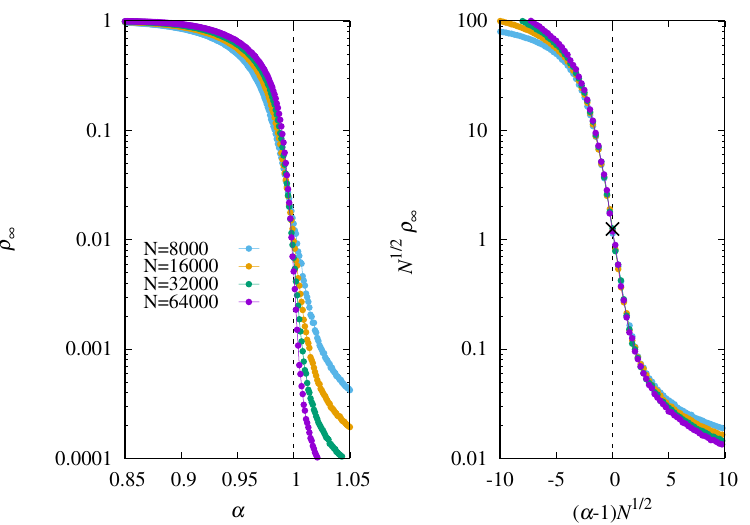}
\caption{Finite-size scaling of the cascade transition for the power function distribution.
(\textbf{Left}) The expected final macroscopic active fraction $\rho_\infty$ as a function of the susceptibility index $\alpha$ for system sizes $N = 8000, 16000, 32000$, and $64000$. The vertical dashed line marks the theoretical transition point at $\alpha = 1$. 
(\textbf{Right}) Data collapse of the macroscopic fractions obtained by plotting the rescaled order parameter $N^{1/2} \rho_\infty $ against the finite-size scaling variable $(\alpha - 1) N^{1/2}$, confirming the critical exponent $\nu = 2$. The vertical dashed line indicates the critical point $(\alpha - 1) N^{1/2} = 0$. The cross symbol ($\times$) at the coordinates $(0, \sqrt{\pi/2})$ represents the exact analytical prediction for the finite-size scaling at $\alpha = 1$. In both panels, the $y$-axis is displayed on a logarithmic scale to highlight the intersection points and the precise scaling behavior. All data points represent Monte Carlo ensemble averages over $10^5$ independent threshold realizations. }
\label{fig:monte_carlo_alpha}
\end{figure}

The left panel of Figure~\ref{fig:monte_carlo_alpha} displays the expected macroscopic order parameter $\rho_\infty$ as a function of the susceptibility index $\alpha$ for progressively larger populations.  As anticipated by the deterministic theory, highly volatile societies ($\alpha \ll 1$) reliably produce system-wide cascades ($\rho_\infty \to 1$), whereas conservative societies ($\alpha > 1$) quickly stifle the instigator, restricting the final size to a negligible microscopic fraction ($\rho_\infty \to 0$). The $y$-axis is plotted on a logarithmic scale to clearly reveal the crossings of the curves for different values of $N$. As the population size increases, these intersection points systematically shift toward $\alpha = 1$ and $\rho_\infty = 0$. The existence of these intersections is a signature of a discontinuous phase transition where $\rho_\infty$ jumps abruptly from $0$ to $1$, a result that perfectly corroborates our deterministic analysis.

At the transition point $\alpha = 1$, we can solve for the cascade propagator factor exactly (see Appendix \ref{appB})  and obtain $Q_k = k^{k-2}/N^{k-1}$. Substituting this  internal propagation factor back into the full probability distribution yields
\begin{equation}
P_N(k) = \binom{N-1}{k-1} \frac{k^{k-2}}{N^{k-1}} \left( 1 - \frac{k}{N} \right)^{N-k}.
\end{equation}
A remarkable  simplification occurs when computing the expectation value for the macroscopic order parameter. By multiplying $P_N(k)$ by $k$, the expression  rearranges into a standard binomial probability evaluated precisely at its own success probability, $p = k/N$,
\begin{equation}
k P_N(k) = \binom{N}{k} \left(\frac{k}{N}\right)^k \left( 1 - \frac{k}{N} \right)^{N-k} \equiv B(k; N, k/N).
\end{equation}
Consequently, the expected active macroscopic fraction $\rho_\infty$ reduces to the sum of these specific binomial terms
\begin{equation}
\rho_\infty =  \frac{1}{N} \sum_{k=1}^N B(k; N, k/N).
\end{equation}
For large population sizes ($N \gg 1$), we can evaluate this sum asymptotically. Applying Stirling's approximation to the binomial coefficient yields $B(k; N, k/N) \approx [2\pi N (k/N) (1 - k/N)]^{-1/2}$. Introducing the continuous intensive variable $x = k/N$, with $dx = 1/N$, the discrete sum transitions directly into the integral
\begin{equation}\label{eq:rho_scaling}
\rho_\infty \approx \frac{1}{\sqrt{2\pi N}} \int_0^1 \frac{dx}{\sqrt{x(1-x)}} =  \sqrt{\frac{\pi}{2N}}.
\end{equation}
This large-$N$ scaling resolves the non-trivial drifting crossover observed in our Monte Carlo simulations. At the critical point $\alpha = 1$, the macroscopic cascade does not survive as an extensive $\mathcal{O}(1)$ fraction of the population, nor does it immediately die. Instead, the final active fraction vanishes sub-linearly, strictly following a $1/\sqrt{N}$ scaling law.

To characterize the finite-size dynamics in the immediate vicinity of the transition point $\alpha = 1$, we employ the standard framework of finite-size scaling \cite{Privman_1990,Landau_2000}. We assume that near the critical point, the macroscopic active fraction follows the  scaling relation 
\begin{equation}
\rho_\infty = N^{-1/2} f\left[ (\alpha - 1) N^{1/\nu} \right],
\label{eq:fss}
\end{equation}
where $f(z)$ is a  scaling function and $\nu$ is the critical exponent that determines the effective width of the transition region \cite{Privman_1990}. To ensure consistency with our exact analytical derivation at the transition point,  Equation (\ref{eq:rho_scaling}), the scaling function must strictly satisfy the boundary condition $f(0) = \sqrt{\pi/2} \approx 1.25$. 

The right panel of Figure~\ref{fig:monte_carlo_alpha} validates this scaling hypothesis. By plotting the rescaled order parameter $N^{1/2} \rho_\infty $ against the rescaled tuning parameter $(\alpha - 1) N^{1/\nu}$, we test the collapse of the Monte Carlo data for different system sizes. Setting the critical exponent to $\nu = 2$ yields a perfect data collapse across all evaluated populations. This precise alignment not only confirms the validity of the scaling ansatz but also establishes that the critical window separating the conservative and volatile regimes narrows proportionally to $N^{-1/2}$.  The critical exponent $\nu = 2$ is derived analytically in Appendix \ref{appC} using a large-deviation analysis.

\section{Intermediate Stable Cascades in Beta-Distributed Thresholds}\label{sec:B}

To better capture realistic scenarios of social polarization, we investigate a regime where radical and conservative factions coexist within the same population. This behavior is naturally modeled by parameterizing individual activation thresholds via a standard Beta distribution. The corresponding cumulative distribution function is defined as $F(\rho) = I_\rho(\alpha, \beta)$, where $I_\rho(\alpha, \beta)$ is the regularized incomplete Beta function \cite{Abramowitz_1972}. In this section, we focus strictly on the regime where the right-tail parameter is bounded below unity ($\beta < 1$), while allowing the left-tail parameter $\alpha$ to vary freely.

Physically, fixing $\beta < 1$ ensures a permanent cluster of high-threshold, conservative agents who strictly resist global activation. Because of this stubborn minority, the system can never achieve full mobilization ($\rho = 1$). In this context, the left-tail parameter $\alpha$ acts as our global tuning variable, regulating the density of hyper-susceptible individuals at the origin. By varying $\alpha$ across the critical point $\alpha = 1$, we can trigger and analyze a clean, continuous phase transition from a completely inactive state to a partially active, intermediately stable cascade.

\subsection{Deterministic Analysis}

The deterministic evolution of the cascade is governed by the discrete-time mapping (\ref{rec_t}), where the long-term steady-state fraction of active nodes corresponds to the fixed points satisfying the transcendental equation 
\begin{equation}\label{fix_p05}
\rho_\infty = I_{\rho_\infty}(\alpha, \beta).
\end{equation}
Because $I_0(\alpha, \beta) = 0$ and $I_1(\alpha, \beta) = 1$ for all valid shape parameters, the boundaries $\rho_\infty = 0$ and $\rho_\infty = 1$ always exist as trivial fixed points of the dynamics. The existence and stability of any intermediate equilibrium $\rho_\infty \in (0, 1)$ depend crucially on the derivative of this regularized incomplete Beta function, whose explicit density was introduced in Equation~(\ref{eq:Beta}). By fixing $\beta < 1$, the exponent $\beta - 1$ is strictly negative, causing the derivative to diverge at the upper boundary, i.e., $\left. dI_\rho / d\rho \right|_{\rho=1} = \infty$. Since this derivative is greater than $1$, the fully mobilized state ($\rho_\infty = 1$) remains unstable across all parameter ranges.

Consequently, the global stability landscape of the system is entirely governed by the behavior of the derivative at the origin ($\rho \to 0$), which exhibits a sharp bifurcation depending on the value of $\alpha$. In the inactive regime where $\alpha > 1$, the positive exponent $\alpha - 1$ forces the derivative at the origin to vanish, $\left. dI_\rho / d\rho \right|_{\rho=0} = 0$. Since this value is strictly less than $1$, the trivial absorbing state $\rho_\infty = 0$ acts as a stable attractor. Concurrently, because the derivative is below unity at the origin ($\rho=0$) and diverges at the upper bound ($\rho=1$), the continuous function $I_\rho(\alpha, \beta)$ cannot intersect the identity line $y = \rho$ inside the open interval $(0,1)$ without violating the single-inflection-point constraint of the distribution, meaning that no intermediate fixed point can exist for $\alpha > 1$.

At the exact critical threshold $\alpha = 1$, the derivative at the origin simplifies to the finite constant $\left. dI_\rho / d\rho \right|_{\rho=0} = 1/B(1, \beta) = \beta$. Since we have explicitly fixed $\beta < 1$, the condition $\left. dI_\rho / d\rho \right|_{\rho=0} < 1$ still holds at this boundary, ensuring that the origin remains locally stable, while signaling that a new stable equilibrium continuously branches out from the origin as $\alpha$ decreases further.

When $\alpha$ drops below unity into the active regime, the exponent $\alpha - 1$ becomes negative, causing the derivative at the origin to diverge, $\left. dI_\rho / d\rho \right|_{\rho=0} = \infty$. Because this derivative is now greater than $1$, the trivial inactive state $\rho_\infty = 0$ becomes strictly unstable. Since both boundaries are now simultaneously unstable ($\left. dI_\rho / d\rho \right|_{\rho=0} > 1$ and $\left. dI_\rho / d\rho \right|_{\rho=1} > 1$), the continuous curve $I_\rho(\alpha, \beta)$ is geometrically forced to cross the identity line from above to below at a unique intersection point inside the open interval. This topological constraint guarantees the existence of a unique, globally stable intermediate fixed point $\rho_\infty \in (0, 1)$.

To characterize the precise nature of the phase transition occurring at $\alpha = 1$, we examine the asymptotic behavior of the order parameter in the joint limit $\alpha \to 1^-$ and $\rho_\infty \to 0^+$. In this regime, the threshold distribution near the origin scales as $\rho_\infty \approx \beta \rho_\infty^\alpha$. Solving this fixed-point relation  for the stable fraction yields the non-analytic scaling law
\begin{equation}\label{exact_beta}
\rho_\infty \approx \exp\left( \frac{\ln(\beta)}{1-\alpha} \right).
\end{equation}
Because $\beta < 1$, the argument inside the exponential remains strictly negative, which implies that the active fraction vanishes faster than any power law as the critical point is approached from below.

Physically, this result reveals that $\alpha = 1$ does not mark a standard Landau-like continuous phase transition \cite{Stanley_1971}, but rather introduces an essential singularity into the order parameter. This behavior is  reminiscent of an infinite-order Berezinskii-Kosterlitz-Thouless (BKT) transition \cite{Berezinskii_1971,Kosterlitz_1973}. Just as in physical systems governed by BKT mechanisms, all derivatives of the order parameter with respect to the tuning variable $\alpha$ vanish identically at the critical point. This essential singularity indicates that while the high density of hyper-susceptible individuals destabilizes the inactive state for any $\alpha < 1$, the resulting intermediate cascade remains drastically suppressed in the immediate vicinity of the critical point, displaying an exceptionally smooth, infinite-order onset before expanding into a robust, partially mobilized equilibrium.

\subsection{Finite Population Analysis}

\begin{figure}[th]
\centering
 \includegraphics[width=1\columnwidth]{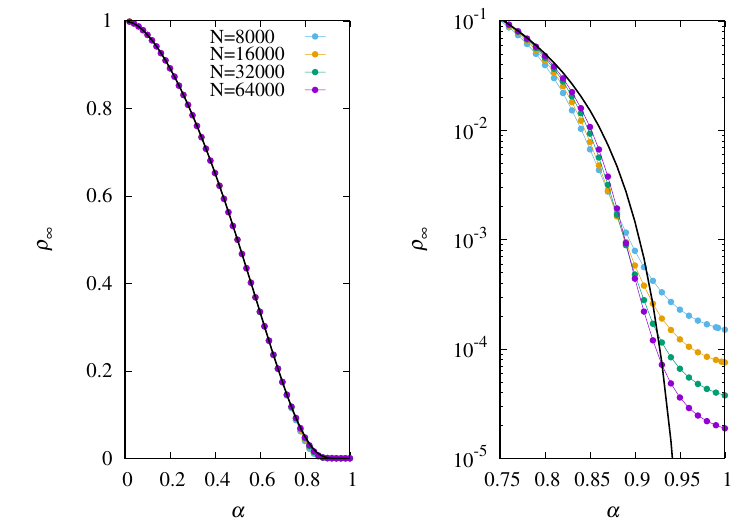}
\caption{Cascade transition equilibrium for a Beta threshold distribution with shape parameter $\beta = 1/2$. (\textbf{Left}) The expected final macroscopic active fraction $\rho_\infty$ as a function of the control parameter $\alpha$ for system sizes $N = 8000, 16000, 32000$, and $64000$. (\textbf{Right}) A logarithmic magnification of the critical region, exposing the pronounced finite-size effects that are otherwise obscured on a linear scale. In both panels, the solid curves represent the infinite population limit obtained from the numerical solution of the fixed-point Equation (\ref{fix_p05}). All data points are Monte Carlo ensemble averages computed over $10^6$ independent threshold realizations.}
\label{fig:monte_carlo_beta05}
\end{figure}

The study of finite-size effects for this general scenario of beta-distributed thresholds presents a significant challenge. Remarkably, away from the critical region, finite-size corrections are strongly suppressed. This is evident in the left panel of Figure \ref{fig:monte_carlo_beta05}, which displays---for the representative case of $\beta = 1/2$---an excellent agreement between the finite-$N$ Monte Carlo simulations and the deterministic infinite population limit obtained by numerically solving Equation (\ref{fix_p05}). 

The resulting finite-size deviations become prominent well before the critical point, tracking back to $\alpha \approx 0.8$, as illustrated in the logarithmic magnification in the right panel of Figure \ref{fig:monte_carlo_beta05}. Far from being negligible, these size-dependent corrections manifest as a massive splitting of the curves, with the effective crossing points between different system sizes occurring around $\alpha \approx 0.85$---strikingly distant from the true critical point at $\alpha = 1$. This significant displacement of the pseudo-critical region is a hallmark of the underlying essential singularity. Furthermore, because the active fraction drops to extreme scales between $10^{-4}$ and $10^{-5}$ in this regime, resolving such minute values against statistical fluctuations required an extensive ensemble of $10^6$ independent realizations.

\begin{figure}[th]
\centering
 \includegraphics[width=1\columnwidth]{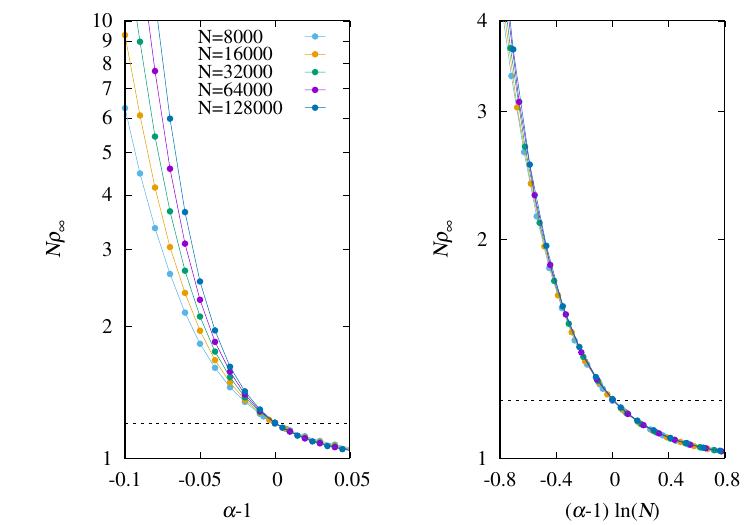}
\caption{Finite-size scaling of the cascade transition for a Beta threshold distribution with shape parameter $\beta = 1/2$. (\textbf{Left}) The expected final macroscopic number of active agents $N \rho_\infty$ as a function of the distance to the critical point $\alpha-1$ for system sizes $N = 8000, 16000, 32000, 64000$, and $128000$. (\textbf{Right}) $N \rho_\infty$ as a function of the scaled distance to the critical point $(\alpha-1) \ln (N)$, showing the collapse of the data for different values of $N$ predicted by the scaling relation (\ref{scale_beta}). The horizontal dashed line indicates $N \rho_\infty \approx 1.20$ at the critical point $\alpha=1$. All data points are Monte Carlo ensemble averages computed over $10^6$ independent threshold realizations.}
\label{fig:monte_carlo_beta05_scaled}
\end{figure}

To characterize the behavior of the order parameter $\rho_\infty$ in the vicinity of the critical point $\alpha = 1$, we employ a finite-size scaling analysis. Crucially, at the exact critical point ($\alpha = 1$), we find that the active fraction scales as $\rho_\infty \approx 1.20/N$. This power-law decay is clearly manifested in the right panel of Figure \ref{fig:monte_carlo_beta05}, where doubling the system size $N$ results in a uniform vertical displacement of the curves, signaling a clean $N^{-1}$ scaling. Furthermore, in the left panel of Figure \ref{fig:monte_carlo_beta05_scaled}, we exhibit a semi-log plot of $N \rho_\infty$ against the distance to the critical point, $\alpha-1$. The perfect collapse of the curves for different values of $N$ at $N \rho_\infty \approx 1.20$ when $\alpha=1$ confirms this scaling behavior, with the inclusion of the larger system size $N=128000$ further validating the asymptotic robustness of the collapse.

Moreover, this collapse remains robust across the entire region $\alpha \ge 1$. As $\alpha$ increases, the number of active individuals $N \rho_\infty$ systematically decreases toward $1$, representing the physical limit where cascades are completely suppressed, leaving only the lone instigator active. In contrast, in the supercritical regime (e.g., $\alpha < 0.75$) where the fraction of active agents becomes independent of $N$, the absolute number of active individuals increases linearly with the system size. A key methodological advantage of adopting $N \rho_\infty$ as the scaling variable is that the complex crossings of the curves observed in the right panel of Figure \ref{fig:monte_carlo_beta05} disappear completely in the left panel of Figure \ref{fig:monte_carlo_beta05_scaled}, yielding a clean, monotonic representation of the transition.

By combining the $\rho_\infty \sim N^{-1}$ scaling at $\alpha=1$ with the exact dependence of the order parameter on the distance to the critical point given by Equation (\ref{exact_beta}), we can formulate a scaling relation that completely characterizes the system in the critical region, 
\begin{equation}\label{scale_beta}
\ln (\rho_\infty) = -\ln (N) g_\beta \left[ (\alpha-1) \ln (N) \right],
\end{equation}
where the scaling function $g_\beta(z)$ satisfies $\lim_{z\to -\infty} g_\beta(z) = \ln(\beta)/z$ to successfully recover the infinite population limit, while $g_\beta(0) = 1 - \ln (A_\beta)/\ln (N)$. Here, $A_\beta$ denotes the expected number of active agents at the exact threshold $\alpha=1$. We recall that $A_{1/2} \approx 1.20$ and that $A_\beta \ge 1$, since the population is always initialized with a single instigator. Furthermore, we find that $A_\beta$ increases monotonically from $1$ as $\beta$ increases, ultimately diverging when $\beta \to 1$. This behavior is physically expected: as $\beta$ decreases toward $0$, the threshold density exhibits a severe power-law divergence at $x=1$, meaning the population becomes overwhelmingly dominated by highly conservative individuals with maximum thresholds. In this limit, secondary triggering is entirely suppressed, forcing $A_\beta$ to its lower bound of $1$. Conversely, as $\beta \to 1$, the population shifts away from this extreme conservatism, allowing larger cascade fluctuations to develop at the critical point, which drives the divergence of $A_\beta$.
The right panel of Figure \ref{fig:monte_carlo_beta05_scaled} displays the remarkable collapse of the $N\rho_\infty$ curves from the left panel when plotted against the scaled distance to the critical point, $(\alpha-1) \ln (N)$, thereby demonstrating the validity of the scaling relation (\ref{scale_beta}). This scaling implies that the  finite-size window around the critical point contracts as $(\ln N)^{-1}$.

The complete analytical foundation for these results is established in Appendix \ref{appD}  using a  tailored large-deviation analysis, which yields the scaling variable $z = (\alpha-1) \ln (N)$ from first principles. Furthermore, by evaluating this framework exactly at $\alpha=1$, we prove that the explicit system-size dependence within the combinatorial prefactor cancels out entirely. Consequently, the intensive macroscopic fraction $\rho_\infty = \langle k \rangle / N$ is shown to scale as $1/N$ at the critical point.

\section{Conclusion}\label{sec:conc}

In this work, we derive an exact analytical solution for the finite-size Granovetter threshold model and identify the finite-size scaling corrections governing the system near two distinct phase transitions. By parameterizing heterogeneity with a Beta threshold distribution with shape parameters $\alpha$ and  $\beta$, we show how variation in intrinsic individual thresholds---capturing agents’ propensity to join a social movement---changes the resulting macroscopic behavior. In the main text, finite-size scaling analysis of Monte Carlo data provides a clear pedagogical account of how finite populations deviate from classical deterministic predictions, while the Appendices develop the exact solution via a large-deviation analysis, offering first-principles verification of the scaling laws.

For the power threshold distribution case ($\beta = 1$), the system exhibits a discontinuous, first-order phase transition at the critical parameter $\alpha = 1$, where the active fraction of agents---those who have adhered to the movement---jumps abruptly from $0$ to $1$, and the finite-size critical region is shown to shrink as $N^{-1/2}$. In stark contrast, when $\beta < 1$, the system undergoes an opposite collective behavior: an infinite-order phase transition at $\alpha = 1$, where the proportion of agents that adhered to the movement increases smoothly from zero, and the critical scaling window contracts much more slowly as $(\ln N)^{-1}$.

For any population characterized by a fixed right-tail parameter $\beta \le 1$, the point $\alpha = 1$ emerges as the universal critical parameter because the underlying threshold density diverges ($\alpha < 1$) or remains finite ($\alpha = 1$) at the origin, structurally guaranteeing a sufficient density of highly susceptible agents to sustain the initial cascade growth. In a social context, the finite-size critical window around $\alpha = 1$ represents the precise domain where collective inertia transitions into widespread mobilization, such as the sudden outbreak of protests \cite{,Chwe_1999}, the viral adoption of technologies \cite{Rogers_1995,Valente_1996,Tilles_2015}, or rapid shifts in public opinion \cite{Galam_2004}. Quantifying the exact width of this critical window is important for systemic interventions: if an organizer wishes to foster a movement, or conversely, if an authority aims to curb an unwanted mobilization within a finite population, they must know precisely how far the parameter $\alpha$ must be shifted to guarantee the stability or suppression of the cascade.

We note that the present framework purposely excludes the parameter regime where $\beta > 1$. In this case, the threshold probability density vanishes at $x = 1$, reflecting a population that lacks highly conservative agents with maximum thresholds. Consequently, while full collective activity ($\rho_\infty = 1$) remains the only asymptotic outcome for $\alpha < 1$, the region where $\alpha > 1$ is characterized by a strict bistability where both $\rho_\infty = 0$ and $\rho_\infty = 1$ constitute stable equilibria. Under these conditions, initializing the dynamics with a lone instigator is insufficient to overcome the critical basin boundary, leaving the system trapped in the inactive state. Activating a macroscopic cascade when $\beta > 1$ and $\alpha > 1$ would therefore require a fundamental alteration of the seeding mechanism, replacing the single instigator with a finite, macroscopic fraction of initial triggers. While this introduces an additional parameter that lies beyond the scope of our current single-seed approach, it uncovers a highly compelling scenario involving competing basins of attraction and nucleation-like phenomena that we reserve for future work.

Looking ahead, a particularly promising avenue for future research involves expanding this finite-size analytical framework to the historically overlooked dimension of multi-group competition. Interestingly, the conceptual seeds for this perspective were planted by Granovetter himself; in a notable footnote of his 1978 paper  \cite{Granovetter_1978}, he explicitly recognized that exploring the temporal evolution of behavior in groups---drawing analogies to James Coleman's mathematical sociology \cite{Coleman_1961}---would require introducing the threshold concept into frameworks that traditionally lacked individual heterogeneity. While this open challenge has been recently addressed from a numerical standpoint by adapting Granovetter's binary paradigm to scenarios of competitive aggregation~\cite{Ishikawa_2026} and casual group formation~\cite{Fontanari_2023,Mariano_2025}, these models still lack closed-form solutions. The exact finite-size probabilistic framework developed in this paper could serve as the mathematical foundation to analytically solve these multi-group environments, shifting the focus from a single movement growing against an inactive background to the exact finite-size scaling of coexisting, rival factions.

\section*{Acknowledgments}

FF is partially supported by  Conselho Nacional de Desenvolvimento Cient\'{\i}fico e Tecnol\'ogico  grant number 305620/2021-5.

\appendix

\section{Asymptotic Derivation of the Propagation Factor $Q_k$}\label{appA}
\renewcommand{\theequation}{A.\arabic{equation}}
\setcounter{equation}{0}
\setcounter{figure}{0}

In our formulation, the internal cascade propagation factor $Q_k$ is implicitly defined by the exact normalization condition (\ref{eq:nQk}), 
\begin{equation}\label{eq:norm_Q}
\sum_{j=1}^{k} \binom{k-1}{j-1} Q_j \left[ 1 - F\left(\frac{j}{N}\right) \right]^{k-j} = 1.
\end{equation}
We wish to derive an  approximation for $Q_j$  in the infinite population  limit ($N \to \infty$). Let $\rho = k/N$ be the macroscopic target size, and $y = j/N$ be the intermediate cascade size, such that $y \in (0, \rho]$. To capture how the propagation factor scales with the system size, we introduce the intensive large-deviation ansatz
\begin{equation}
Q_j \sim \exp \left [ N \psi(y) \right ],
\end{equation}
where the intensive rate function $\psi(y)$  is formally defined as the scaling limit
\begin{equation}
\psi(y) \equiv \lim_{N \to \infty} \frac{1}{N} \ln Q_{\lfloor Ny \rfloor}.
\end{equation}
This function represents the exponential decay rate (or entropic penalty) associated with the cascade successfully propagating up to a continuous fraction $y$ of the population.

Using Stirling's approximation ($\ln n! \approx n \ln n - n$), the binomial coefficient for large $N$ transforms into an intensive entropy-like expression
\begin{equation}
\ln \binom{k-1}{j-1} \approx \ln \binom{N\rho}{Ny} \approx N \Big[ \rho \ln \rho - y \ln y - (\rho - y) \ln (\rho - y) \Big].
\end{equation}
Substituting this scaling into Equation (\ref{eq:norm_Q}), the discrete summation transitions into a continuous integral dominated by an exponential weight
\begin{equation}
1 \approx N \int_0^\rho \exp\Big\{ N \, S(y; \rho) \Big\} dy,
\label{eq:laplace_int}
\end{equation}
where the action $S(y; \rho)$ is defined as
\begin{equation}
S(y; \rho) = \rho \ln \rho - y \ln y - (\rho - y) \ln (\rho - y) + \psi(y) + (\rho - y) \ln \left[ 1 - F(y) \right].
\end{equation}

According to Laplace's method for asymptotic integrals, the integral in Equation (\ref{eq:laplace_int}) is entirely dominated by the value of the integrand at its global maximum, $y^*$. For the probability to be perfectly conserved---evaluating to exactly $1 = e^0$ for any arbitrarily large $N$---the maximum value of the action must vanish identically for \textit{any} physically reachable target density $\rho$
\begin{equation}
S(y^*(\rho); \rho) = 0.
\label{eq:zero_max}
\end{equation}
Since this zero-maximum identity must hold across all choices of $\rho$, the total derivative of the action with respect to $\rho$ along the saddle-point trajectory must be strictly zero, 
\begin{equation}
\frac{d S}{d \rho} = \frac{\partial S}{\partial \rho} + \frac{\partial S}{\partial y} \frac{d y^*}{d \rho} = 0.
\end{equation}
Because $y^*$ is an interior maximum, the stationarity condition requires the partial derivative with respect to the internal state to vanish ($\partial S / \partial y = 0$). We are therefore left with the envelope condition
\begin{equation}
\frac{\partial S}{\partial \rho} \Bigg|_{y=y^*} = 0.
\end{equation}
Computing this partial derivative explicitly yields
\begin{equation}
\frac{\partial S}{\partial \rho} = \ln \rho + 1 - \ln(\rho - y) - 1 + \ln\left[ 1 - F(y) \right] = \ln\left( \frac{\rho}{\rho - y} \right) + \ln\left[ 1 - F(y) \right].
\end{equation}
Equating this expression to zero at the saddle point $y = y^*$ gives
\begin{equation}
\ln\left( \frac{\rho [1 - F(y^*)]}{\rho - y^*} \right) = 0, 
\end{equation}
which implies
\begin{equation}
y^* = \rho F(y^*).
\label{eq:saddle_point}
\end{equation}
Now, we enforce the zero-maximum constraint from Equation (\ref{eq:zero_max}) to solve explicitly for the unknown rate function
\begin{equation}
\psi(y^*) = -\rho \ln \rho + y^* \ln y^* + (\rho - y^*) \ln (\rho - y^*) - (\rho - y^*) \ln \left[ 1 - F(y^*) \right].
\end{equation}
Substituting  our saddle-point trajectory (\ref{eq:saddle_point})  into this expression yields
\begin{equation}
\psi(y^*) = y^* \ln\left(\frac{y^*}{\rho}\right)  = y^* \ln F(y^*).
\end{equation}
Because this relation must hold for any valid intermediate state of the cascade process, we obtain the intensive rate function constructively as $\psi(y) = y \ln F(y)$.  Exponentiating  the rate function recovers the propagation factor  in the infinite population limit 
\begin{equation}
Q_k \approx \exp\Big\{ N \psi(k/N) \Big\} = \exp\Big\{ k \ln F(k/N) \Big\} = \left[ F\left(\frac{k}{N}\right) \right]^k.
\end{equation}

\section{Exact Solution for the Uniform Distribution Baseline ($\alpha = 1$)}\label{appB}
\renewcommand{\theequation}{B\arabic{equation}}
\setcounter{equation}{0}

For a uniformly distributed population of thresholds ($F(x) = x$), the recurrence relation for the cascade propagation factor, given by Equation~(\ref{eq:Qk}) in the main text, reduces to
\begin{equation}
Q_k = 1 - \sum_{j=1}^{k-1} \binom{k-1}{j-1} Q_j \left( 1 - \frac{j}{N} \right)^{k-j},
\label{eq:app_Qk}
\end{equation}
with the base case $Q_1 = 1$. To motivate the general functional form of $Q_k$, we explicitly evaluate the first few cascade sizes directly from Equation~(\ref{eq:app_Qk}):
\begin{align}
Q_1 &= 1, \nonumber \\
Q_2 &= 1 - \binom{1}{0} Q_1 \left(1 - \frac{1}{N}\right) = 1 - \left(1 - \frac{1}{N}\right) = \frac{1}{N} = \frac{2^{2-2}}{N^{2-1}}, \nonumber \\
Q_3 &= 1 - \left[ \binom{2}{0} Q_1 \left(1 - \frac{1}{N}\right)^2 + \binom{2}{1} Q_2 \left(1 - \frac{2}{N}\right) \right] \nonumber \\
    &= 1 - \left[ \left(1 - \frac{2}{N} + \frac{1}{N^2}\right) + 2\left(\frac{1}{N}\right)\left(1 - \frac{2}{N}\right) \right] = \frac{3}{N^2} = \frac{3^{3-2}}{N^{3-1}}.
\end{align}
From the emerging structural pattern of these initial terms, we naturally infer the general ansatz
\begin{equation}
Q_j = \frac{j^{j-2}}{N^{j-1}}.
\label{eq:app_ansatz}
\end{equation}
To prove that this inferred relation holds for all $k \ge 1$, we deploy strong mathematical induction. We assume as our induction hypothesis that Equation~(\ref{eq:app_ansatz}) is valid for all intermediate sizes $j \in \{1, 2, \dots, k-1\}$. Substituting this hypothesis into the sum in Equation~(\ref{eq:app_Qk}) yields
\begin{equation}
Q_k = 1 - \sum_{j=1}^{k-1} \binom{k-1}{j-1} \left[ \frac{j^{j-2}}{N^{j-1}} \right] \frac{(N - j)^{k-j}}{N^{k-j}} = 1 - \frac{1}{N^{k-1}} \sum_{j=1}^{k-1} \binom{k-1}{j-1} j^{j-2} (N - j)^{k-j}.
\label{eq:app_pre_abel}
\end{equation}
The remaining algebraic summation can be evaluated exactly by invoking Abel's binomial identity \cite{Riordan_1968},  which states that
\begin{equation}
\sum_{m=0}^{n} \binom{n}{m} (x + m)^{m-1} (y + n - m)^{n-m} = \frac{1}{x} (x + y + n)^n.
\label{eq:abel_identity}
\end{equation}
By setting $n = k - 1$ and shifting the summation index via $m = j - 1$, the sum in Equation~(\ref{eq:app_pre_abel}) represents a truncated Abel sum with parameters $x = 1$ and $y = N - k$, missing only its final boundary term ($j=k$). Evaluating this complete sum and subtracting the missing $j=k$ configuration gives
\begin{equation}
\sum_{j=1}^{k-1} \binom{k-1}{j-1} j^{j-2} (N - j)^{k-j} = \underbrace{N^{k-1}}_{\text{Full Abel Sum}} - \underbrace{\binom{k-1}{k-1} k^{k-2} (N-k)^0}_{j=k \text{ boundary term}} = N^{k-1} - k^{k-2}.
\end{equation}

Finally, inserting this result back into Equation~(\ref{eq:app_pre_abel}) closes the induction step,
\begin{equation}
Q_k = 1 - \frac{1}{N^{k-1}} \left( N^{k-1} - k^{k-2} \right) = 1 - 1 + \frac{k^{k-2}}{N^{k-1}} = \frac{k^{k-2}}{N^{k-1}}.
\end{equation}
This completes the proof that $Q_k = k^{k-2}/N^{k-1}$ is the exact analytical solution for the uniform cascade baseline.

It is instructive to verify that this exact finite-size solution perfectly recovers the asymptotic macroscopic behavior derived in Appendix \ref{appA}.  By factoring the exact expression, we obtain:
\begin{equation}
Q_k = \frac{k^{k-2}}{N^{k-1}} = \left( \frac{k}{N} \right)^k \left( \frac{N}{k^2} \right).
\end{equation}
The first term corresponds to the extensive exponential weight $[F(k/N)]^k$ predicted by Laplace's method for the uniform distribution $F(x) = x$. The second algebraic term, $N/k^2$, represents a sub-exponential finite-size correction arising naturally from fluctuations around the saddle point. 

To formally demonstrate that this correction vanishes in the infinite population  limit, we evaluate the intensive rate function $\psi(x)$ for the macroscopic fraction $x = k/N$:
\begin{eqnarray}
\psi(x) &\equiv& \lim_{N \to \infty} \frac{1}{N} \ln Q_{\lfloor Nx \rfloor} \nonumber \\
        &=& \lim_{N \to \infty} \left[ \frac{Nx}{N} \ln\left(\frac{Nx}{N}\right) + \frac{\ln N - 2 \ln(Nx)}{N} \right] \nonumber \\
        &=& \lim_{N \to \infty} \left[ x \ln x - \frac{\ln N}{N} - \frac{2 \ln x}{N} \right].
\end{eqnarray}
Because the logarithmic finite-size terms strictly vanish as $N \to \infty$, the rate function  converges to $\psi(x) = x \ln x$. This confirms that our exact baseline solution is  consistent with the general asymptotic scaling  $Q_k \approx \exp\{N x \ln F(x)\}$.

\section{Large-Deviation  Analysis for the Power Distribution ($\beta = 1$)}\label{appC}
\renewcommand{\theequation}{C\arabic{equation}}
\setcounter{equation}{0}

To analytically determine the critical exponent $\nu$ that governs the width of the finite-size transition window for the power function distribution, we examine the behavior of the system in the immediate vicinity of the critical point by setting $\alpha = 1 + \epsilon$, where $\epsilon \ll 1$.

As shown in Section \ref{sec:fin}, the exact probability distribution for a cascade of size $k$ in a system of total size $N$ is given by
\begin{equation}
P_N(k) = \binom{N-1}{k-1} Q_k \left[ 1 - F\left(\frac{k}{N}\right) \right]^{N-k},
\end{equation}
where $Q_k$ is the internal cascade propagation factor and $F(x)$ is the cumulative threshold distribution evaluated at the intensive scale $x = k/N$.

From the asymptotic analysis derived in Appendix \ref{appA}, the leading-order exponential scaling of the propagation factor is given by $Q_k \approx [F(x)]^k$. Substituting this into the exact expression for $P_N(k)$ and using the combinatorial identity $\binom{N-1}{k-1} = x \binom{N}{k}$, we isolate the structural terms dependent on the extensive system size $N$. For large population sizes ($N \gg 1$), we apply Stirling's approximation to the binomial coefficient $\binom{N}{k}$. Tracking only the dominant exponential terms, the probability distribution can be expressed in the standard large-deviation form \cite{Touchette_2009}
\begin{equation}\label{eq:C2}
P_N(k) \sim \mathcal{A}(N, x) \exp\left[ N \cdot \Phi(x) \right],
\end{equation}
where $\mathcal{A}(N, x)$ represents a non-exponential, algebraic prefactor, and $\Phi(x)$ is the intensive scaling function that dictates the shape of the probability landscape
\begin{equation}\label{eq:C3}
\Phi(x) = x \ln \left( \frac{F(x)}{x} \right) + (1-x) \ln \left( \frac{1 - F(x)}{1 - x} \right).
\end{equation}
Because the critical exponent $\nu$ uniquely governs the width of the finite-size transition window, its value is determined exclusively by the curvature and scaling of the exponential argument $\Phi(x)$ near the critical point, remaining  invariant to the explicit form of the algebraic prefactor $\mathcal{A}(N, x)$.

Evaluating the behavior of the scaling function $\Phi(x)$ near the critical point by substituting the cumulative power function distribution, $F(x) = x^\alpha = x^{1+\epsilon}$, and expanding to leading order in $\epsilon$ yields
\begin{equation}
\Phi(x) \approx - \epsilon^2 \frac{x^2 \ln^2 x}{2(1 - x)}.
\end{equation}
Reinserting this result into the large-deviation expression shows that the exponential envelope of the probability distribution scales as
\begin{equation}
P_N(k) \sim \mathcal{A}(N, x) \exp \left[ - N \epsilon^2 \frac{x^2 \ln^2 x}{2(1 - x)} \right].
\end{equation}
The emergence of the coupled product $N\epsilon^2$ in the exponential governing factor implies that the critical fluctuations are controlled by the invariant combination $\epsilon N^{1/2}$. Consequently, the finite-size critical scaling window contracts as $N^{-1/2}$, which directly establishes the correlation length exponent as $\nu = 2$.

\section{Large-Deviation  Analysis for the Beta Distribution ($\beta < 1$)}\label{appD}
\renewcommand{\theequation}{D\arabic{equation}}
\setcounter{equation}{0}

To establish the analytical origin of the logarithmic scaling variable $z = (\alpha -1)  \ln(N)$ in the  case of the Beta Distribution  with shape parameter $\beta < 1$, we  use the general large-deviation expressions for $P_N(k)$ and the exponential argument $\Phi(x)$ given in Equations (\ref{eq:C2})  and (\ref{eq:C3}) of Appendix \ref{appC}.  As before,  we  evaluate the system's behavior in the immediate vicinity of the critical point by setting $\alpha = 1 + \epsilon$, with $\epsilon \ll 1$.

To evaluate $\Phi(x)$ for the Beta distribution, we first examine the behavior of the cumulative threshold distribution $F(x) = \frac{1}{B(\alpha, \beta)} \int_0^x t^{\alpha-1}(1-t)^{\beta-1} dt$ near the origin ($x \ll 1$). Expanding the term $(1-t)^{\beta-1} \approx 1$ to leading order, the integration yields
\begin{equation}
F(x) \approx \frac{x^\alpha}{\alpha B(\alpha, \beta)} = \frac{x^{1+\epsilon}}{(1+\epsilon) B(1+\epsilon, \beta)}.
\end{equation}
In the critical limit where $\epsilon \to 0$, the normalization factor reduces  to $1/B(1, \beta) = \beta$. Therefore, the cumulative distribution close to the critical  point scales as
\begin{equation}
F(x) \approx \beta x^{1+\epsilon}.
\end{equation}

We now substitute this result into the intensive scaling function $\Phi(x)$. The first term in Equation  (\ref{eq:C3}) becomes
\begin{equation}\label{eq:D3}
x \ln \left( \frac{F(x)}{x} \right) \approx x\ln \left( \beta x^\epsilon \right) =x  \ln \beta + \epsilon x \ln x.\end{equation}
For the second term, we expand the logarithms for $x \ll 1$, 
\begin{equation}\label{eq:D4}
(1-x) \ln \left( \frac{1 - \beta x^{1+\epsilon}}{1 - x} \right) \approx (1-x) \left[ -\beta x^{1+\epsilon} - (-x) \right] \approx x - \beta x^{1+\epsilon}.
\end{equation}
To explicitly isolate the role of the critical perturbation $\epsilon$, we expand the power-law term as $x^{1+\epsilon} = x e^{\epsilon \ln x} \approx x(1 + \epsilon \ln x)$. Substituting this into Equation (\ref{eq:D4})  gives
\begin{equation}
x - \beta x^{1+\epsilon} \approx x(1 - \beta) - \beta \epsilon x \ln x.
\end{equation}
Combining the pieces from Equations (\ref{eq:D3}) and (\ref{eq:D4})  into the full expression for $\Phi(x)$, given in Equation  (\ref{eq:C3}),  we find
\begin{equation}
\Phi(x) \approx x \ln \beta + \epsilon x \ln x + x(1 - \beta) - \beta \epsilon x \ln x,
\end{equation}
which simplifies to
\begin{equation}
\Phi(x) \approx x(1 - \beta + \ln \beta) + (1 - \beta)\epsilon x \ln x.
\end{equation}
Crucially, unlike the pure power-law distribution (where $\beta = 1$) analyzed in Appendix \ref{appC}, the first-order linear term in $\epsilon$ does not vanish when $\beta < 1$.

To map this result onto the extensive system size, we evaluate the full exponential envelope $N \cdot \Phi(x)$ by returning to the discrete cascade scale via the substitution $x = k/N$,
\begin{equation}
N \cdot \Phi\left(\frac{k}{N}\right) \approx k(1 - \beta + \ln \beta) + (1 - \beta)\epsilon k \ln\left(\frac{k}{N}\right).
\end{equation}
Expanding the logarithm of the ratio yields $\ln(k/N) = \ln k - \ln N$. In the limit $N \gg 1$ for sub-macroscopic cascades ($k \ll N$), the term $\ln N$ asymptotically dominates over $\ln k$. Retaining the leading finite-size scaling contribution alongside the intrinsic local threshold terms, the exponent reduces to
\begin{equation}
N \cdot \Phi\left(\frac{k}{N}\right) \approx k(1 - \beta + \ln \beta) - (1 - \beta) k \cdot \left[ \epsilon \ln N \right].
\end{equation}
Consequently, the probability distribution of small cascades takes the asymptotic form
\begin{equation}\label{eq:D10}
P_N(k) \sim \mathcal{A}(N, k/N) \cdot e^{k(1 - \beta + \ln \beta)} \cdot \exp\left[ -k (1 - \beta) (\alpha-1) \ln N  \right].
\end{equation}
Notice that as $\beta \to 1$, the coefficient $(1-\beta)$ vanishes, forcing the system out of this log-scaling regime and requiring the  higher-order expansion used in Appendix \ref{appC}.

Finally, evaluating $P_N(k)$ exactly at the critical point ($\alpha = 1$, or equivalently $\epsilon = 0$) allows us to recover the microscopic scaling regime.  In this case, the argument of the second exponential in Equation~(\ref{eq:D10}) vanishes, reducing the extensive large-deviation argument to
\begin{equation}
N \cdot \Phi\left(\frac{k}{N}\right) \approx k(1 - \beta + \ln \beta).
\end{equation}
Because the functional combination $1 - \beta + \ln \beta$ is strictly negative for all shape parameters satisfying $\beta < 1$, we can define a positive characteristic damping factor $\sigma_\beta = |1 - \beta + \ln \beta| > 0$. Consequently, the critical probability distribution for the absolute cascade size $k$ can be written as
\begin{equation}\label{eq:D11}
P_N(k) \sim \mathcal{A}(N, k/N) \exp\left( -k \sigma_\beta \right).
\end{equation}
To evaluate the finite-size scaling of the prefactor in the sub-macroscopic limit ($k \ll N$), we note from the Stirling expansion of the combinatorial kernel that the structural system-size dependence enters via a factor of $\sqrt{N x}$ in the denominator. Upon substituting the discrete cascade scale $x = k/N$, this term reduces to $\sqrt{k}$, effectively canceling the explicit role of $N$. Therefore, in the limit $N \gg 1$, the prefactor loses its explicit scaling dependence on the system size, becoming an $N$-independent local function $\mathcal{A}(N, k/N) \approx \mathcal{A}(k)$.

Due to this lack of $N$-dependence in both the prefactor and the exponential envelope, the expected absolute number of active agents at the critical point converges to a finite, constant value determined exclusively by the local cascade combinatorics and the intrinsic threshold geometry, $\langle k \rangle = \sum_{k} k P_N(k) \equiv A_\beta$. It follows directly that the intensive, macroscopic fraction of active individuals $\rho_\infty = \langle k \rangle / N$ obeys the  scaling relation
\begin{equation}
\rho_\infty = \frac{A_\beta}{N} \sim \frac{1}{N},
\end{equation}
analytically reproducing the finite-size behavior obtained from the numerical Monte Carlo simulations.

\end{document}